\documentclass[12pt,tightenlines,eqsecnum,floats,shownopacs,nofootinbib,amsmath,amssymb,aps,prd]{revtex4}
\begin{document}

\title{Viewpoint: Simplicity of Black Holes}
\author{Abhay Ashtekar}
 \affiliation{Institute for Gravitation and the
Cosmos \& Physics Department, Penn State, University Park, PA 16802,
U.S.A.}

\begin{abstract}

The Editors of \emph{Viewpoint} requested a brief report on Norman G\"urlebeck's recent results \cite{1} concerning black holes in astrophysical environments. The first draft I wrote assumed knowledge of general relativity at the level of a graduate-level course. But it was significantly simplified to make the final version \cite{viewpoint} accessible to undergraduates who are not familiar with general relativity. This submission to arXiv contains the first version that is likely to be more helpful to beginning researchers who are interested in black holes. 

\end{abstract}

\pacs{}

\maketitle

In general relativity, the gravitational field is encoded in the very geometry of space-time. One of the most dramatic consequences is that the geometry can be warped into configurations that we interpret as black holes. But this encoding comes at a cost: in place of a single, linear, Poisson equation of Newtonian gravity, we now have a set of complicated, non-linear Einstein equations. Yet, not only can one obtain exact solutions representing isolated black holes in equilibrium, but these solutions are characterized just by two numbers, the mass $M$ and the angular momentum $J$ \cite{heusler}! On learning this astonishing fact, S. Chandrasekhar described his experience as ``the most shattering realization'' of his entire scientific life \cite{chandra}. 

A Physical Review Letter by G\"urlbeck \cite{1} unravels a new aspect of this simplicity of black holes. The paper restricts itself to non-rotating black holes in equilibrium, i.e., \emph{static} black holes. If a static black hole is isolated, it's geometry is described by the Schwarzschild solution which is completely characterized just by its mass $M$. But astrophysical black holes are rarely isolated; they are surrounded by discs and magnetic fields. Since the black hole uniqueness results require not only equilibrium but also \emph{source-free} Einstein's equations, now the space-time geometry is no longer described by the Schwarzschild metric but is distorted by matter. Because of the underlying non-linearity, the distortions can be quite complicated. But if the entire system is static, the distortions can be neatly encoded in a set of `mass multipole moments' $M_{n}$ in a coordinate invariant fashion. 

To understand why this is possible, let us first consider Newtonian gravity which is governed by the Poisson equation ${\Delta} \phi = 4\pi\, G\, \rho$. One can define the \emph{field multipoles} by carrying out an expansion of the Newtonian potential $\phi$ in powers of $1/r$. As is well-known (e.g., from electrostatics) the Poisson equation enables one to recast these numbers as \emph{source multipoles} obtained by integrating the matter density $\rho$, weighted by suitable powers of $r$ and spherical harmonics. In static space-times of general relativity, using the norm of the time translation symmetry vector field, one can introduce the analog $\tilde\phi$ of the Newtonian potential $\phi$. Einstein's equations imply that it must satisfy the Laplace equation ${\tilde\Delta} \tilde\phi = 0$ outside sources. Using the asymptotic behavior of $\tilde\phi$, Geroch \cite{geroch} introduced a set of field  multipoles that completely characterize the space-time geometry in the entire source-free region \cite{beig}. For the Schwarzschild metric representing an isolated black hole, only the monopole --the mass-- is non-zero while for a generic black hole we have the entire tower of multipoles which capture the distortions in the space-time geometry outside sources. The presence of higher multipoles encapsulates the fact that the space-time geometry of realistic, astrophysical black holes can be quite complicated. In the terminology introduced by John Wheeler, such black holes \emph{can} have `hair' in general relativity.

While the origin of Geroch's multipoles has close similarity with the origin of field multipoles in Newtonian gravity, there are also important differences. In general relativity, the system of equations is non-linear because now the Laplacian $\tilde\Delta$ refers to the curved spatial metric which is itself determined by $\tilde\phi$. As a consequence, it has not been possible to relate Geroch's field multipoles to sources. Consider a black hole which is in equilibrium only in the sense that there is no flux of matter and energy across its horizon, although there may be (even non-static) matter and radiation away from it. In this case, the geometry of the \emph{black hole horizon} would not be spherically symmetric. If it is axisymmetric, these distortions in its geometry can be captured completely in a set of invariantly defined, horizon multipoles, even though the horizon is in a highly non-linear, strong field regime \cite{ih}. They represent \emph{source multipoles} of black holes in a precise sense. But they do not equal Geroch's field multipoles, defined at infinity, even in globally static solutions. This is because the field multipoles, defined at infinity, receive contributions not only from the black hole but also from the outside matter, \emph{as well as} the gravitational field itself,  because gravity acts as its own source through non-linearities of general relativity.

However, as Weyl pointed out almost a century ago \cite{weyl}, a key simplification occurs if one assumes that the entire space-time is not only static but also axisymmetric. Now one can replace $\tilde\phi$ by another field $U$ which satisfies the Laplace equation ${\Delta} U =0$ outside sources, where the Laplacian now refers to a \emph{fiducial, flat} spatial metric. To obtain a black hole solution to Einstein's equations, one can first solve the Laplace equation in \emph{flat space} with suitable boundary conditions to ensure the existence of a horizon, and then determine all the metric coefficients from the solution $U$ \cite{gh}. Geroch's multipoles $M_{n}$ can be read off from the coefficients $U_{n}$ in the asymptotic expansion of $U$ along z-axis, defined by the rotational symmetry \cite{fhp}. Thus, in effect, non-linearities of Einstein's equations can be ironed out in Weyl solutions. Using this fact, in an earlier paper \cite{gurlebeck}, G\"urlebeck showed that it is possible to express each coefficient $U_{n}$ as a sum of two surface integrals, one on a 2-sphere $S_{H}$ that encloses only the black hole horizon, and another on a 2-sphere $S_{M}$ surrounding just the matter in the axisymmetric discs. To summarize, thanks to the underlying linearity of Weyl solutions, the field moments defined at infinity can be expressed as a sum of integrals that can be attributed to individual source components, just as in the Newtonian theory. 

G\"urlebeck's new and surprising observation is that the integrals over $S_{H}$ representing the black hole contributions to Geroch's \emph{field multopoles} are non-zero \emph{only for the monopole}! All the higher field moments (as well as a part of the monopole) can be attributed entirely to matter. Thus, even in the astrophysical situations where the horizon is highly distorted because of external matter rings, the black hole contributes only to the mass and not to any of the `hair' \emph{seen at infinity}. In this sense, even black holes with distorted horizons have an unforeseen simplicity, at least if they are not rotating.

This result suggests interesting directions for future research. On the conceptual side, the challenge is to understand the relation between the source multipoles that characterize the distortions in the horizon geometry \cite{ih}, and the integrals on $S_{H}$ which `dress' them appropriately to provide the black hole contribution to field multipoles at infinity. How does this `dressing' manage to hide all the \emph{horizon hair} form a distant observer?  On the mathematical side, the key question is whether the results can be extended to rotating black holes, where Einstein's equations \emph{cannot} be cast in an effectively liner form \`a la Weyl. Interestingly, analysis in \cite{gurlebeck} was carried out using certain inverse scattering methods which \emph{do} extend to rotating, axisymmetric black holes. However, it is not clear if, at the end, the horizon multipoles will be again be so dressed as to be hidden for the distant observers. Will they now see only the Kerr field multipoles \cite{hansen}? Finally, recently there has been considerable interest in observationally testing the no-hair theorems (see, e.g., \cite{tests}). Since these theorems refer to general relativity and need not hold in alternate theories of gravity, observations could constrain alternate theories and provide new tests of general relativity. However, at the simplest level, such tests require clean candidates, since the no-hair theorems refer to source-free Einstein's equations. If there is sufficient matter to distort the horizon geometry, one needs to first determine whether a given observational mission measures source (i.e., horizon) multipoles, or the field multipoles at infinity. If it is the field multipoles, results of \cite{1} could be used to test general relativity. However, there are still two caveats: (i) the black hole and the matter rings have to be approximately static and axisymmetric; and, (i) the observational mission should also be able to measure the matter multipoles encoded in the integral over $S_{M}$ sufficiently accurately to test if the observed higher multipoles can be completely attributed to these integrals. This makes the task rather difficult. Nonetheless, as observational missions expand their reach, it is interesting to have yet another avenue to test the no hair theorem for black holes in `dirty' astrophysical environments .

\medskip
This work was supported in part by the NSF grant PHY-1205388 and the Eberly Research Funds of Penn State.

\end{document}